\documentclass[journal]{IEEEtran}
\usepackage[cmex10]{amsmath}
\usepackage{flushend}
\usepackage{cite}
\usepackage{url}
\usepackage{ifpdf}
\ifCLASSINFOpdf
  \usepackage[pdftex]{graphicx}
  \DeclareGraphicsExtensions{.pdf}
\else
  \usepackage[dvips]{graphicx}
   \DeclareGraphicsExtensions{.eps}
\fi
\usepackage{graphicx}
\usepackage{amssymb}
\usepackage{epstopdf}
\usepackage{amsmath}
\usepackage{multirow}
\usepackage{algorithm}
\usepackage{algpseudocode}

\graphicspath{{figures/}}

\newcommand{\bk}{{\mathbf k}}

\newcommand{\cP}{{\mathcal P}}

\DeclareMathOperator*{\argmax}{arg\,max}

\begin{document}
\title{Data Driven SMART Intercontinental Overlay Networks}

\author{Olivier Brun\thanks{O. Brun is with LAAS/CNRS,
7 Avenue du Colonel Roche, 31077 Toulouse Cedex 4, France.},  
Lan Wang and Erol~Gelenbe \IEEEmembership{Life~Fellow,~IEEE}
\thanks{L. Wang and E. Gelenbe are with the Department
of Electrical and Electronic Engineering, Imperial College, London SW7 2BT, UK.}
\thanks{Manuscript received August 25, 2015; revised XX XX, 201Y.}}
\maketitle
\begin{abstract}
This paper addresses the use of Big Data and machine learning based analytics to the real-time
management of Internet scale Quality-of-Service Route Optimisation with the help of an overlay network.
Based on the collection of large amounts of data sampled each $2$ minutes over a large number of source-destinations pairs, we show that intercontinental Internet Protocol (IP) paths are far from optimal with respect to 
Quality of Service (QoS) metrics such as end-to-end round-trip delay. We therefore develop a machine learning based scheme that exploits large scale data collected from communicating node pairs in a multi-hop overlay network that uses IP between the
overlay nodes themselves, to select paths that provide substantially better
QoS than IP. The approach inspired from Cognitive Packet Network protocol, uses Random Neural Networks with Reinforcement Learning based on the massive data that is collected, to select intermediate overlay hops resulting in significantly better QoS than IP itself. The routing scheme is illustrated on a $20$-node intercontinental overlay network that collects close to $2\times 10^6$ measurements per week, and makes scalable distributed routing decisions. Experimental results show that this approach improves QoS significantly and efficiently in a scalable manner.\end{abstract}

\maketitle

\section{Introduction}
Autonomic communications \cite{Dobson} were introduced as a means to set-up and adaptively manage large scale networks based on user needs, without direct human intervention. Although the field emerged from active networks \cite{Galis1}, it is making its mark in software overlay networks \cite{Galis2} and distributed system design \cite{Swesi}. In the future, autonomic communications may further increase their hold through the flexibility offered by Software Defined Networks (SDN) \cite{Chemouil}.

Well known network measurements have shown that IP (Internet Protocol) routing often results in paths that are sub-optimal with respect to a number of metrics~\cite{Savage:1999, Paxson1996}. Besides, measurements have also established  that the routing scalability of the Internet comes at the expense of reduced fault-tolerance of end-to-end communications between Internet hosts~\cite{Labovitz1998, LabovitzBGP, Dahlin01end-to-endwan,Han2004,Paxson1996}. Current routing protocols may work reasonably well when only ``best effort'' delivery is required, but the requirements for modern distributed services are typically far more stringent, demanding greater performance and availability of end-to-end routes than these protocols can deliver. The ideal solution would be a complete rethink of the Internet routing infrastructure, doing away with the existing architecture and redesigning it with the benefit of hind-sight about its deficiencies. Unfortunately, the Internet has become resistant to major changes, preventing even necessary changes to take place. 

\begin{figure}[!htb]
  \centering
    \includegraphics[width=0.9\columnwidth]{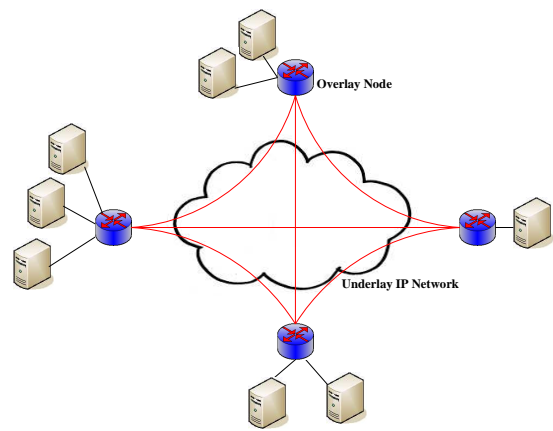}
    \caption{Schematic description of the structure of a routing overlay, where the Overlay Nodes exchange packets with each other with packets that tunnel through
    the IP connections, while paths between Overlay Nodes may transit through intermediate Overlay Nodes.}
    \label{fig:overlay-networks}
  \end{figure}

On the other hand, routing overlays have been proposed as a method for improving performance, without the need to re-engineer the underlying network~\cite{Peterson04,Touch03,Feamster04,Beck03}. The basic idea is to move some of the control over routing into the hands of end-systems. As illustrated in Figure \ref{fig:overlay-networks}, a routing overlay is formed by software routers, which are deployed in different spots over the Internet. The overlay nodes monitor the quality of the Internet routes between themselves and  cooperate with each other to share data.  By adding intermediate routing hops into the path taken by streams of packets, they influence the overall path taken by the packets, without modifying the underlying IP mechanism for computing routes. In a routing overlay, the endpoints of the information exchange are unchanged from what they would have been in the absence of the overlay, but the route through the network that the packets traverse may be quite different.

Routing overlays can be used to quickly recover from path outages, and also  improve the QoS of data flows. Indeed, the overlay nodes constantly monitor the IP routes between themselves so that failed parts of the Internet can be avoided when a node detects that the primary Internet path is subject to anomalies. Similarly, this approach makes it possible to override the routes determined by Internet protocols and to route traffic based on metrics directly related to the performance needs of the application.

The Resilient Overlay Network (RON) \cite{Andersen01resilientoverlay} was the first routing overlay to be implemented in a wide-area network, demonstrating that adding an extra (overlay) routing hop could benefit an application in terms of improved delay and reachability. To find and use alternate paths, RON monitors the health of the underlying Internet paths between overlay nodes, dynamically selecting paths that avoid faulty areas. The main drawback of RON is that it does not scale very well: as the number of participating routers $N$ increases, the $O(N^2)$ probing overhead becomes a limiting factor, because RON uses all-pairs probing. The downside is that a reasonable RON overlay can support only about $50$ routers before the probing overhead becomes overwhelming. However RON has inspired many other approaches \cite{Gummadi2004,Hu2004,Nakao2006}, but no existing work has tackled the problem of building an overlay that can be widely and efficiently deployed over a sizable population of routers. 

In this paper, we investigate the use of the \emph{Cognitive Packet Network} (CPN)  \cite{Gelenbe2004} to design and evaluate a {\em scalable routing overlay}. In previous work, CPN has been shown to be effective for a variety of uses, including QoS optimisation, network security enhancement and energy savings \cite{Sakellari}, and adaptive routing of evacuees in emergency situations \cite{DBES}. The overlay we design is based on a set of proxies installed at different Cloud servers, or they may be in other servers across the network, so that the overlay itself operates over these proxies from some source to destination in source routed manner, while the flow of packets between proxies travels in conventional IP mode. The routing between proxies provides QoS-driven source routing, and performs self-improvement in a distributed manner by learning from QoS measurements gathered on-line, and discovering new routes \cite{Gelenbe2009}. 

This data driven intercontinental packet routing scheme constantly, say every two minutes,  collects round-trip delay data at the overlay nodes; it then makes scalable distributed decisions using a machine learning approach from this massive amount of data. In view of the $N$ overlay nodes used in our experiments, every two minutes the system may collect up to $N^2$ data points. Thus over $24$ hours with $20$ overlay nodes, each checking connectivity and round-trip-delays (RTT) with $19$ other nodes, the network can collect up to some $2.7\times 10^5$ data points per day. 
However, our work shows that most of the benefit of the technique is achieved when only a small number of alternate paths are tested, so that there can be considerable reduction in complexity of data processing and decision making. Furthermore, it is also possible to use the full CPN scheme for the overlay, which means that only the best one-overlay-hop connections are probed, as when CPN seeks the paths with the best QoS across a large network.

The routing decisions are made on-line at each proxy of the overlay network based on adaptive learning techniques using the random neural network (RNN) \cite{Gelenbe1993}. Each overlay node
uses a RNN which  is trained using Reinforcement Learning with the data collected at the node itself, while intermediate IP routers proceed using standard Internet routing. 

The rest of this paper is organized as follows. Section \ref{sec:routing-overlay} describes SMART, our self-healing and self-optimizing routing overlay. Section \ref{sec:learning-with-RNN} is devoted to the adaptive learning techniques we use in SMART, based on the random neural network in order to learn the optimal routes in the overlay with a modest monitoring effort. Section \ref{sec:results} presents the experimental results we obtain with an intercontinental overlay network, and Section \ref{sec:conclusion} draws some conclusions and suggests further work.


\section{SMART: a self-healing and self-optimizing routing overlay}
\label{sec:routing-overlay}
SMART\footnote{Self-MAnaging Routing overlay} is a self-healing, self-optimizing and highly scalable routing overlay that accepts customized routing policies formulated by distributed applications according to their own needs. The overlay network is formed by software routers deployed over the Internet, and it operates by monitoring the quality of Internet paths (latency, bandwidth, loss rate) between overlay nodes and re-routing packets along an alternate path when the primary path becomes unavailable or suffers from congestion.

\begin{figure}[!htb]
  \centering
    \includegraphics[width=0.9\columnwidth]{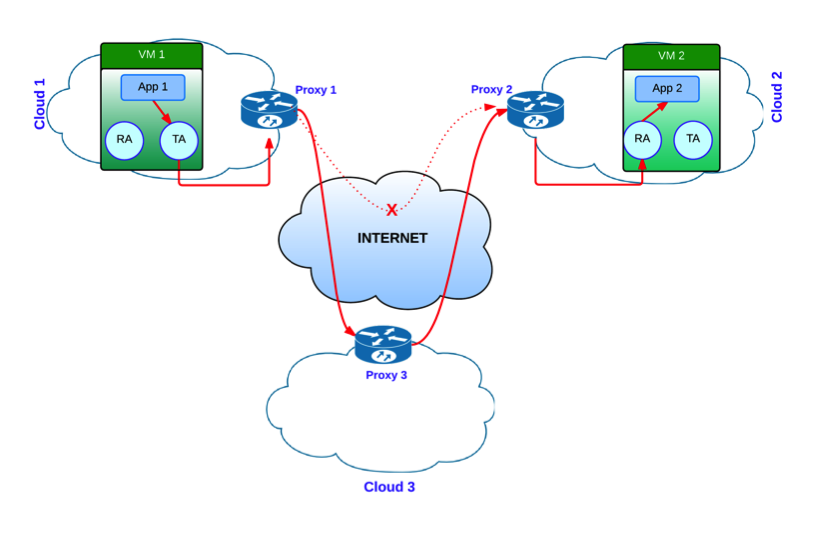}
    \caption{Architecture of the Autonomic Communication Overlay: the Overlay Nodes used by SMART exchange packet streams via the Internet using the Internet Protocol (IP) either directly,
    or via intermediate Overlay Nodes. When a SMART path uses multiple Overlay Nodes, IP is used between adjacent Overlay Nodes.}
    \label{fig:SMART-arch}
  \end{figure}

\begin{figure}[!htb]
  \centering
    \includegraphics[width=0.65\columnwidth]{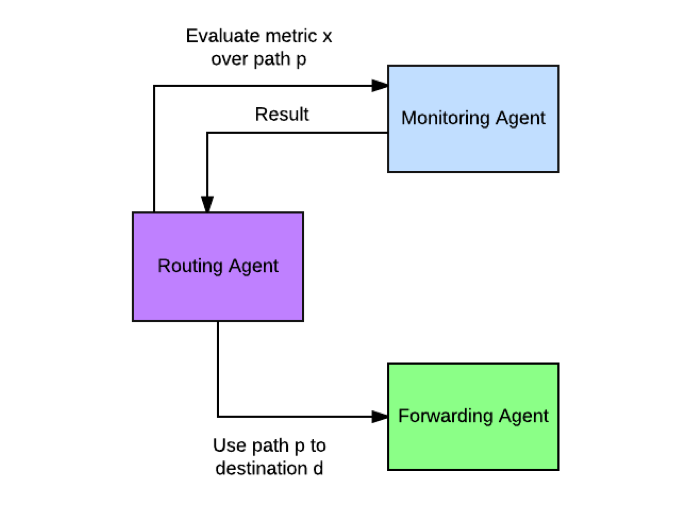}
    \caption{Interactions between the entities constituting the proxy.}
    \label{fig:Proxy-entities}
  \end{figure}

As shown in Figure \ref{fig:SMART-arch}, the SMART overlay network is formed by software agents that are deployed at Virtual Machines (VM) in Cloud sites,  and possibly at other servers
connected to the Internet: 
\begin{itemize}
\item On each VM, a {\bf Transmission (TA) and Reception  Agent} run together with various Applications or tasks. The Proxy, on the other
hand acts as the front-end between the VM and the overlay network.
\item Each VM's software router is the {\bf Proxy} that monitors the quality of the overlay paths towards other destinations, selects the best paths, and forwards the packets over these paths. 
\item The TA  receives the packets that are being sent to other Applications at other sites by an Application running in the same VM.  The TA forwards the packets to the local Proxy using IP-in-IP encapsulation. The Proxy will then handle the forwarding of the packet towards the destination Application a some other VM. 
\item The RA receives packets from the local Proxy, de-enpasulates and delivers them to the appropriate Application in the VM. The TA, RA and Proxy allow us to control the path of 
the packets through the network, without the applications being aware that their data flows are routed by the overlay.
\end{itemize}
The Proxy is the interface for packets into the VM, but it also acts as an intermediate software  router for the overlay as described in Figure \ref{fig:Proxy-entities}.
It is constituted of three software agents:
\begin{itemize}
\item {\bf The Monitoring Agent}: monitors the quality of the Internet paths between the local cloud and the other clouds in terms of latency, bandwidth, and loss rate. The MA can be queried by the routing agent in order to discover the quality of a path according to a given metric, and can be configured to monitor the availability of a path at regular $2$ minute intervals. 
\item {\bf The Routing Agent}: drives the monitoring agent and uses the data it collects to discover an optimal path (e.g., low-latency, high-throughput, etc.) with minimum monitoring effort, and
writes the optimal path for a given destination into the routing table of the forwarding agent.
\item {\bf The Forwarding Agent}: forwards each incoming packet to its destination on the path it was instructed to use by the RA. We use source routing, that is, the routing table of the source proxy describes the complete path {\em between overlay nodes proxies} to be followed by a packet to reach its destination, while the path between proxies is determined by the
conventional IP protocol. Each subsequent Proxy  determines the next Proxy from the information contained in the SMART header. The final Proxy forwards the packet to the appropriate RA, that delivers it to the destination Application.
\end{itemize}

\begin{figure}[!htb]
  \centering
    \includegraphics[width=0.95\columnwidth]{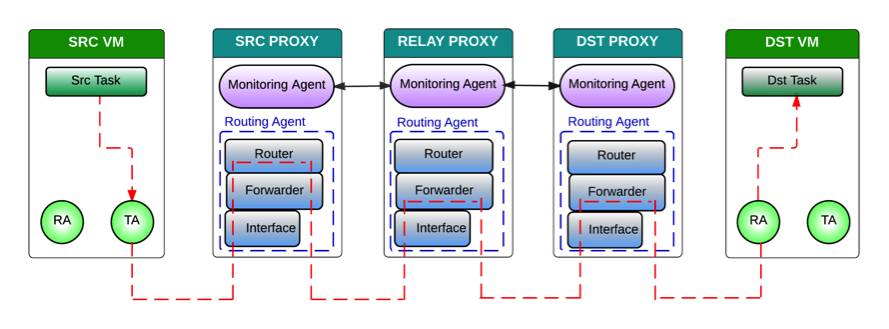}
    \caption{Details of the SMART packet forwarding process.}
    \label{fig:SMART-forwarding}
  \end{figure}

In this architecture, the routing/forwarding of a packet proceeds as shown in Figure \ref{fig:SMART-forwarding}. When a packet is sent by a source task to a destination task located in a different cloud, it is first intercepted  and forwarded to the TA. The TA uses IP-in-IP encapsulation to forward an altered packet to the Proxy. The payload of the altered packet is the original packet, plus an additional SMART header. Upon reception of the SMART packet, the routing agent of the Proxy looks-up its routing table in order to determine the path to the destination.  The sequence of intermediate proxies is written in the SMART header, and then the SMART packet is forwarded to the first one of these proxies. Each intermediate proxy then forwards the packet to the next hop on the path, until the final proxy is reached. When this happens, the packet is forwarded to the RA of the destination VM. The RA de-encapsulates the SMART packet and forwards the original IP packet to the destination task using a raw socket.
 
The optimal path to a destination cloud is found using active monitoring. The monitoring agent of a proxy regularly measures the latency  to the destination cloud by sending probe packets along paths to that destination. Each probe packet is time-stamped by each intermediate node on its way forward and on its way back, so that the source proxy can easily deduce the latency of each segment of the path. Note that the size of the overlay network we are using, and the measurements at $2$ minute intervals imply that $2.7\times 10^5$ data points are being collected per day.

\section{Learning with the Random Neural Network}
\label{sec:learning-with-RNN}

We wish to build a routing overlay that can be widely deployed over a large population of routers, implying that the monitoring effort (that is, the number of probed links per time slot) should grow at most linearly with the number of nodes of the overlay. Instead of requiring a performance guarantee at each time step, we look for an online decision algorithm that uses a limited monitoring effort but achieves asymptotically the same average (per round) end-to-end latency as the best path. The idea is to design an algorithm that exploits past observations so as to quickly learn path performance and efficiently select the optimal path. This can be viewed as a
a multi-armed bandit problem \cite{Cesa-Bianchi2006,Gyorgy2007} in which decisions correspond to choosing paths between the source and the destination. At each successive time slot, the algorithm chooses a subset of paths to probe, and measures the sum of edge delays in the probed paths. The algorithm then sends its packet over the minimum latency path among those it has probed. Delays may change from one time slot to the next one, and our goal is to probe those paths at each time slot, whose total delay incurred by packets traveling over the paths over time is not significantly more than that of the best route from the source to the destination. In other words, probing does not cover {\em all} possible paths but only a few paths which have been observed in previous probing steps
to provide low overall forwarding delay for packets. However, as in the previously tested CPN routing scheme \cite{Gelenbe2009}, we have to widen our probing at random
over other paths, so that we do not miss out on paths whose latency has substantially improved over recent history, and we use Reinforcement Learning \cite{Sutton} to adjust the parameters of a Random Neural Network (RNN), acting as an adaptive critic, as first suggested in \cite{Halici}. The RNN has been used in many other applications, such as image processing and virtual reality \cite{Simulation}.
\subsection{The Random Neural Network}
\label{subsec:RNN}
The random neural network (RNN)  \cite{Gelenbe1993,G-Nets}  is a recurrent model, i.e. it can contain feedback loops as in the present work,  but it can also be used in feedforward mode as with conventional Artificial Neural Networks. It has
a finite number of $n$ interconnected neurons. Its state is a vector $\bk(t) = [\bk_1(t), \bk_2(t), \ldots, \bk_n(t)]$, where $\bk_i(t)$ is a non-negative integer valued {\em random variable} representing the ``potential'' of the $i$-th neuron at time $t$. The probability of the state of the RNN is denoted by $p(k, t) \equiv Pr[\bk(t) = k]$, where $k=(k_1,~...~k_n)$ and the $k_i\geq 0$ are integers.
Its stationary probability distribution function, when it exists,  is $p(k) \equiv \lim_{t \rightarrow \infty} p(k,t)$.

A neuron $i$ of the RNN is said to be excited whenever $\bk_i(t)>0$, in which case it can  fire and send signals at an average rate $r_i$, with exponential, independent and identically distributed inter-spike intervals. Spikes will go from neuron $i$ to neuron $j$ with probability $p_{i, j}^+$ as excitatory spikes, and with probability  $p_{i, j}^-$ as  inhibitory spikes, where
$\sum_{j=1}^n p_{i, j}^+ + p_{i, j}^- +d_i= 1$, $\quad i=1, \ldots, n$,
and $d_i$ is the probability that the fired spike is lost in the network or that it leaves the network towards some external system.

Let $w_{i,j}^+ = r(i) p_{i, j}^+\geq 0$ and $w_{i,j}^- = r(i) p_{i, j}^-\geq 0$; they denote the emission rates of excitation and inhibition signals from neuron $i$ to neuron $j$. 
In addition, for any neuron $i$, exogenous excitatory, and inhibitory spikes, can enter the  neuron from outside the network at Poisson rates
denoted by $\Lambda_i$ and $\lambda_i$, respectively.
By its no-linearity, the mathematical structure of a RNN differs from that of widely used queueing systems such as the Jackson network or the BCMP model \cite{Muntz,Timotheou}.

However, despite this major difference (and the non-existence of properties such as quasi-reversibility of the RNN model equations), it has been shown to have a product form solution
\cite{Gelenbe1993} given by:
\begin{equation}
p(k) = \prod_{i=1}^n (1-q_i) \, q_i^{k_i}, ~q_i = \frac{\lambda_i^+}{r(i)+\lambda_i^-},
\label{eq:product-form}
\end{equation}
where $q_i\lim_{t\rightarrow\infty}Pr[\bk_i(t)>0]$  is the stationary probability that neuron $i$ is excited, 
and $\lambda_i^+$ and $\lambda_i^-$, represent the total flows of excitatory and inhibitory spikes arriving at
neuron $i$, and satisfy a system of nonlinear simultaneous equations:
\begin{equation*}
\lambda_i^+ = \sum_j q_{j} w_{j,i}^+ + \Lambda_i,~\lambda_i^- = \sum_j q_{j} w_{j,i}^- + \lambda_i. 
\end{equation*}
and $q_i$ is obtained from the solution of the following system of non-linear equations:
\begin{equation}
q_i = \frac{\lambda_i^+}{r_i+\lambda_i^-}.\label{eq:lambda} 
\end{equation}
Since it has been proved in \cite{Gelenbe1993} that these equations have an unique solution with all $q_i\in [0,1]^n$, the non-linear system \eqref{eq:lambda} can be solved efficiently using the simple fixed-point iteration:
\begin{equation}
q_i^{k+1}\leftarrow min\big[1,\frac{\sum_j q_{j}^k w_{j,i}^+ + \Lambda_i}{r_i+\sum_j q_{j}^k w_{j,i}^- + \lambda_i}\big],
\end{equation}
starting with the initial values $q_i^0=0.5$ for all the $i=1,~..~,n$.

\subsection{Reinforcement Learning}
\label{subsec:reinforcement-learning}
In the following, we assume that there is a single origin/destination pair and describe the algorithm implemented by the source Proxy for learning an optimal route to the destination Proxy. This algorithm is implemented by the Routing Agent of the Proxy and is based on a RNN. As input, we are given a set $\cP$ of $N$ possible paths to the destination Proxy. Each neuron of the RNN is associated to one of these paths. At regular time intervals, the Routing Agent uses the RNN to select $K$ paths to the destination, monitors the quality of these paths, and then chooses the path with the best performance as described in Algorithm \ref{alg:learning-RNN}. While the approach described in \cite{Gelenbe2004} uses both loss and delay to select paths, here we only focus on delay or latency.

\begin{algorithm}
\caption{Learning optimal paths with a RNN and Reinforcement Learning.}
\label{alg:learning-RNN}
\begin{algorithmic}[1]
\For{$tl=1,~2,\ldots$}
	\State $\cP(t_l)$ is the set of $K$ neurons with highest probabilities $q_i$ at time $t_l$.
	\State $R_j(t_l) \gets$ reward obtained with path $j$.
	\State $p^* \gets \argmax_{j \in \cP(t_l)}R_j(t_l)$
	\For{$j \in \cP(t_l)$}
	\State $\nu_j \gets R_j(t_l)/T(t_l)$.
	\If {$R_j(t_l) \geq T(t_l)$}
		\For{$i=1,\ldots,N$}
	    		\State $\Delta_i \gets \left ( \nu_j - 1\right ) \, w_{i,j}^+$.
    			\State $w_{i,j}^+ \gets w_{i,j}^+ + \Delta_i$.
			\State $w_{i,k}^- \gets w_{i,k}^- + \frac{\Delta_i}{N-2}$, $\forall k \neq j$.
		\EndFor
	\Else
		\For{$i=1,\ldots,N$}
    			\State $\Delta_i \gets \left ( 1-\nu_j \right ) \, w_{i,j}^-$.
    			\State $w_{i,j}^- \gets w_{i,j}^- + \Delta_i$.
			\State $w_{i,k}^+ \gets w_{i,k}^+ + \frac{\Delta_i}{N-2}$, $\forall k \neq j$.
		\EndFor
	\EndIf
	\For{$i=1,\ldots,N$}
		\State $r_i^* = \sum_{k=1}^n w_{i,k}^+ + w_{i,k}^-$.
		\State $w_{i,j}^+ \gets w_{i,j}^+ \, \frac{r_i}{r_i^*}$.
		\State $w_{i,j}^- \gets w_{i,j}^- \, \frac{r_i}{r_i^*}$.
	\EndFor
	\State Solve the non-linear system \eqref{eq:lambda} 
	\State $T(t_{l+1}) \gets \beta \, T(t_l) + (1 - \beta) \, R_j(t_l)$.
	\EndFor
\EndFor
\end{algorithmic}
\end{algorithm}

The function to be minimized is the  routing ``goal'' $G$, in this case the round-trip delay to the destination, which can be measured. Its inverse is the reward $R = G^{-1}$. Let $\cP(t) \subset \cP$ be the set of paths selected at time $t$ by the algorithm. Each of these paths corresponds to one of the neurons with the highest probabilities $q_i$ at time $t$ (line $1$). The algorithm monitors the quality of these paths at instants $t_l$ for successive integers $l$, from which it deduces the reward $R_j(t_l)$ for each path $j \in \cP(t_l)$ (line $2$). For reasons of scalability, the CPN algorithm which is designed for an arbitrarily large network, takes the routing decisions at the single node level, only choosing only the next hop for the ``smart packets'' (SPs) that seek the paths  \cite{Gelenbe2009}, based on the final destination of the SP.

However the algorithm we use for overlay routing deals with relatively small overlay networks with a small number of alternate paths for a given source to destination pair.
Thus it considers each of the monitored full paths.  $R_j(t_l)$  is used to adjust the values of the two matrices $W^+$ and $W^-$ based on Reinforcement Learning, based on the decision threshold $T_{t_l}$:
$$
T(t_l) = \beta \, T(t_{i-l}) + (1 - \beta) \, R_j(t_l),
$$
where $\beta \in (0,1)$ is a real number that is used to introduce forgetfulness: a large $\beta$ will give more importance to recent events. When round-trip delay is the goal function, $T(t_i)$ is the ``exponential average'' up to time $t_i$ of the round-trip delay for the packet in that flow.
To start the learning, we first determine whether the most recent value of the reward $R(t_l)$ is larger than the threshold $T(t_l)$. If that is the case, then we increase significantly the excitatory weights going into neuron $j$ (line $10$), that was the previous winner, rewarding it for its previous success, and increase the inhibitory weights leading to other neurons, but in a much smaller proportion (line $11$). Otherwise, if the new reward is smaller than the previously observed threshold, we increase significantly the inhibitory weights leading to the previous winning neuron (line $16$), punishing it for not being successful last time, and increase in a much smaller proportion all excitatory weights leading to other neurons (line $17$) to give other decisions a greater chance of being selected. 
As in CPN \cite{Gelenbe2009}, to avoid ever-increasing values of the weights, we re-normalize their values (lines $20$-$23$) after each update:
\begin{equation*}
w_{i,j}^+ \gets  w_{i,j}^+, \frac{r_i}{r_i^*},~w_{i,j}^-  \gets  w_{i,j}^- \, \frac{r_i}{r_i^*},
\end{equation*}
where 
$$
r_i^* = \sum_{k=1}^n w_{i,k}^+ + w_{i,k}^-
$$
We then compute the stationary probability $q_i$ that neuron $i$ is excited (line $25$) by solving the following system of nonlinear simultaneous equations
\begin{eqnarray*}
q_i & = & \frac{\lambda_i^+}{r(i)+\lambda_i^-}, \\
\lambda_i^+ & = & \sum_j q_{j} w_{j,i}^+ + \Lambda_i, \\ 
\lambda_i^- & = & \sum_j q_{j} w_{j,i}^- + \lambda_i,
\end{eqnarray*}
as described in Section \ref{subsec:RNN}. Finally, we update the decision threshold $T(t_l)$ (line $26$).

\section{Experimental Results}
\label{sec:results}
\begin{figure}[!htb]
  \centering
    \includegraphics[width=0.8\columnwidth]{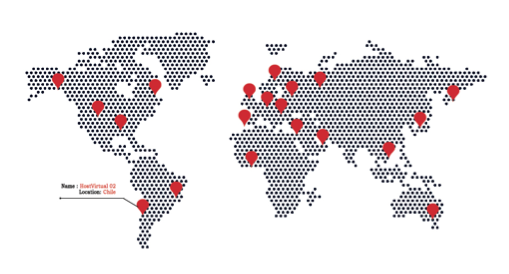}
    \caption{Geographical location of the 20 Source and Destination Nodes used in our experiments within the NLNog ring. These same nodes are also used for our
    direct IP based routing measurements, and also to evaluate the SMART Overlay Node based routing scheme.}
    \label{fig:nlnog-ring-node-loc}
  \end{figure}

 We now describe the results that we obtained with the proposed algorithm during an Internet-scale experiment, where we used  $20$ {\em Overlay Nodes} of the $NLNog$ ring  shown in Figure \ref{fig:nlnog-ring-node-loc}. Note that these overlay nodes are interconnected by literally hundreds of Internet nodes which are unknown to us or the overlay, and which support the overlay itself. 
 
 We first measured the latency and loss rates between all pairs of nodes every two minutes, communicating through the Internet, for a period of one week using the ICMP-based ping utility. Furthermore, when five consecutive packets were lost between a specific pair of nodes, we considered that the particular source was disconnected from that destination.  The path latency was measured as the round-trip time (RTD), i.e. {\em  the length of time it takes 
for a packet to be sent to its destination, plus the length of time it 
takes for the corresponding acknowledgment (ACK) packet to be received at the source}.

\begin{figure}[!htb]
  \centering
    \includegraphics[width=0.9\columnwidth]{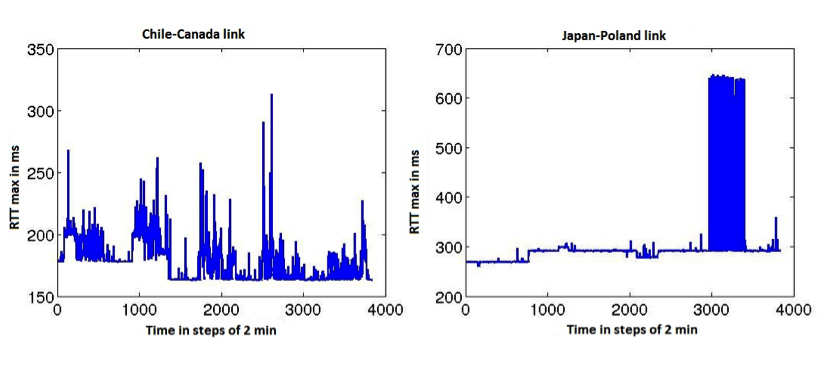}
    \caption{The RTD in milliseconds using the IP protocol, given in averages over successive 2 minute intervals over an observation period; each unit along the x-axis corresponds to successive 2 minute intervals
    over the long time period being shown. The connections measured are between Chile-Canada (left) and Japan-Poland (right), and we observe the large variation in average RTD depending
    on the time period being considered. }
    \label{fig:RTD_nlnog_ring}
  \end{figure}

\begin{figure}[!htb]
  \centering
    \includegraphics[width=0.8\columnwidth]{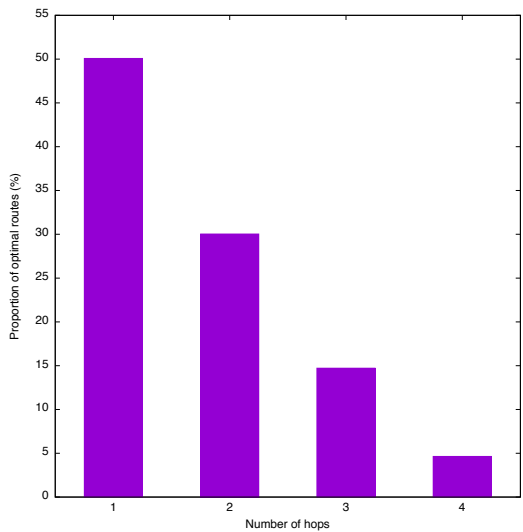}
    \caption{Percentage of instances when the overlay path that minimises RTD, which uses  IP paths between Overlay Nodes, is observed to include  $1$, $2$, $3$ or $4$ Overlay Hops. We see that at most
    two overlay hops cover most of the optimal cases.}
    \label{fig:CDF-nb-hops-NLNog}
  \end{figure}

\begin{figure}[!htb]
  \centering
    \includegraphics[width=0.8\columnwidth]{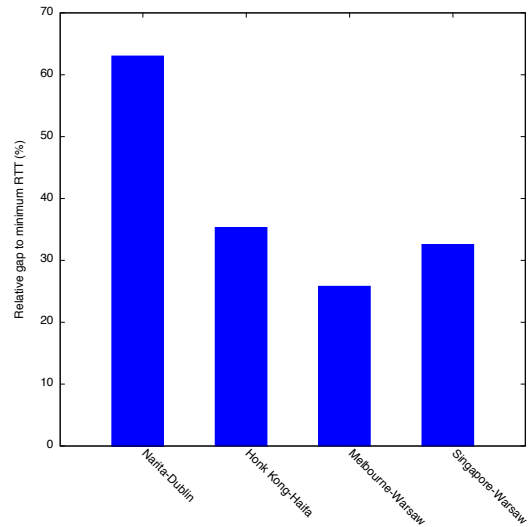}
    \caption{Percentage of average RTD, for the best 2-hop path relative to the minimum RTD path, for certain origin to destination pairs, where RTD is averaged over the successive 2 minute measurement intervals.}
    \label{fig:relative-gap-to-2hops-paths}
  \end{figure}

The main observations from this Internet-scale experiment based on some $2.7\times 10^5$ measurements per day, that was carried out over a whole week
(i.e. a total of roughly $2\times 10^6$ measurements), were the following:
\begin{itemize}
\item There was a path outage across the Internet at least once in the week for 65\% of origin-destination pairs, and 21\% of these path outages lasted more than $4$ minutes. In fact, 11\% of the outages lasted more than $14$ minutes.
\item We observed that the RTD of purely IP routes (without overlays) exhibits strong and unpredictable variations: see Figure \ref{fig:RTD_nlnog_ring}, and these
variations can be as large as  500\%. 
\item Throughout our experiments, the IP route was {\em very clearly} not the minimum latency path in 50\% of the cases, as shown in Figure \ref{fig:CDF-nb-hops-NLNog}. 
\item There was always at least one origin to destination pair whose latency could  be reduced by more than 76\% by selecting an alternate path to the IP route, by using one or more overlay nodes.
\begin{itemize}
\item Surprisingly enough, as shown in Figure \ref{fig:CDF-nb-hops-NLNog}, for 30\% of the cases the optimal path hadonly 2 Overlay-Hops. This shows that a limited deviation from IP can actually produce much better QoS than IP itself.
 \item Then, a natural question is whether it is enough to consider routes of at most two Overlay-Hops.  
 \item Interestingly, in 20\% of the cases the optimal path was a 3 or 4 Overlay-Hop path. 
 \item However, on average, the relative difference in measured delay between the optimum path and  the best 2 Overlay-Hop path was only 3.38\%. 
 \item However, as shown in Figure \ref{fig:relative-gap-to-2hops-paths}, for some origin to destination pairs, there was a significant benefit in using a path which had more than two overlay hops.
 \end{itemize}
\item Similarly, more than 11\% of the IP routes were observed to have a loss rate greater than 1\%. An in particular, by selecting paths via the overlay that {\em had a different path than those proposed by IP}, it was possible to have no packet loss at all.
\end{itemize}
These results have also shown that path outages are routine events in the Internet. Furthermore, the paths selected by IP routing are strongly suboptimal. This Internet-scale experiment confirmed observations similar to those made in previous studies. 

As we will now see, the CPN or RNN based algorithm that exploits Overlay Nodes allows a significant improvement in QoS, more specifically a significant decrease in round-trip delay, with a very modest monitoring and computational effort.

\section{An experiment in decreasing the Round-Trip Delay (RTD)}

In the next experiment, we simplify the routing scheme so that we only consider the direct IP route and 2 Overlay-Hop routes. Therefore the number of possible overlay paths between an Overlay  Source (OS) and Overlay Destination (OD) is $N=19$. Furthermore for a given connection, the algorithm chooses two alternate overlay paths to monitor, and selects the one with the minimum latency. 

Thus from a given source node, the algorithm will measure at most $4$ links per measurement round. The resulting observed average delays are summarized in Table \ref{tab:rnn-avg-results}.

\begin{table}
\begin{center}
\begin{tabular}{|c|c|c|}
\hline
 	                                  & Direct	& 2-hop Overlays \\
	                                  \hline
Percentage of non-optimal instances	& 50.08\%	& 16.20\%  \\
Average percentage difference above minimum latency 	& 11.1\%	 & 4.24\% \\
\hline
\end{tabular}
\caption{Non-Optimal RTDs for all the measurements: IP vs. SMART routing with 2-hop overlays.
16\% of 2-hop overlays and 50\% of IP paths are non-optimal. RTDs for IP can substantially exceed the minima and averages over all measurements.}
\label{tab:rnn-avg-results}
\end{center}
\end{table}

These average values do not truly measure the gains obtained in the pathological routing situations we seek to improve. Thus,  in Table \ref{tab:rnn-pathological-cases}~we present some examples for which the system provides 
significant gains, despite the partial coverage of the overlay topology.

\begin{table}
\begin{center}
\begin{tabular}{|c|c|c|}
\hline
 	                                  & Direct IP 	& K=2 \\
\hline
Singapore-Israel	 & 26.86	& 0.34 \\
Japan-Chile	&  60.73	&  0.08 \\ 
Australia-Chile	&  26.03	&  0.30 \\
Norway-Singapore &	23.35 & 1.15 \\ 
Poland-Brazil	&  24.32    &  0.39 \\
Ireland-Moscow & 119.39	&   0.18 \\ 
Israel-Moscow	&  48.39	&   0.17 \\ 
\hline
\end{tabular}
\caption{Relative gap in percentage of RTD, to the minimum observed RTD, for some pathological origin to destination pairs.}
\label{tab:rnn-pathological-cases}
\end{center}
\end{table}

\begin{figure}[!htb]
  \centering
    \includegraphics[width=0.8\columnwidth]{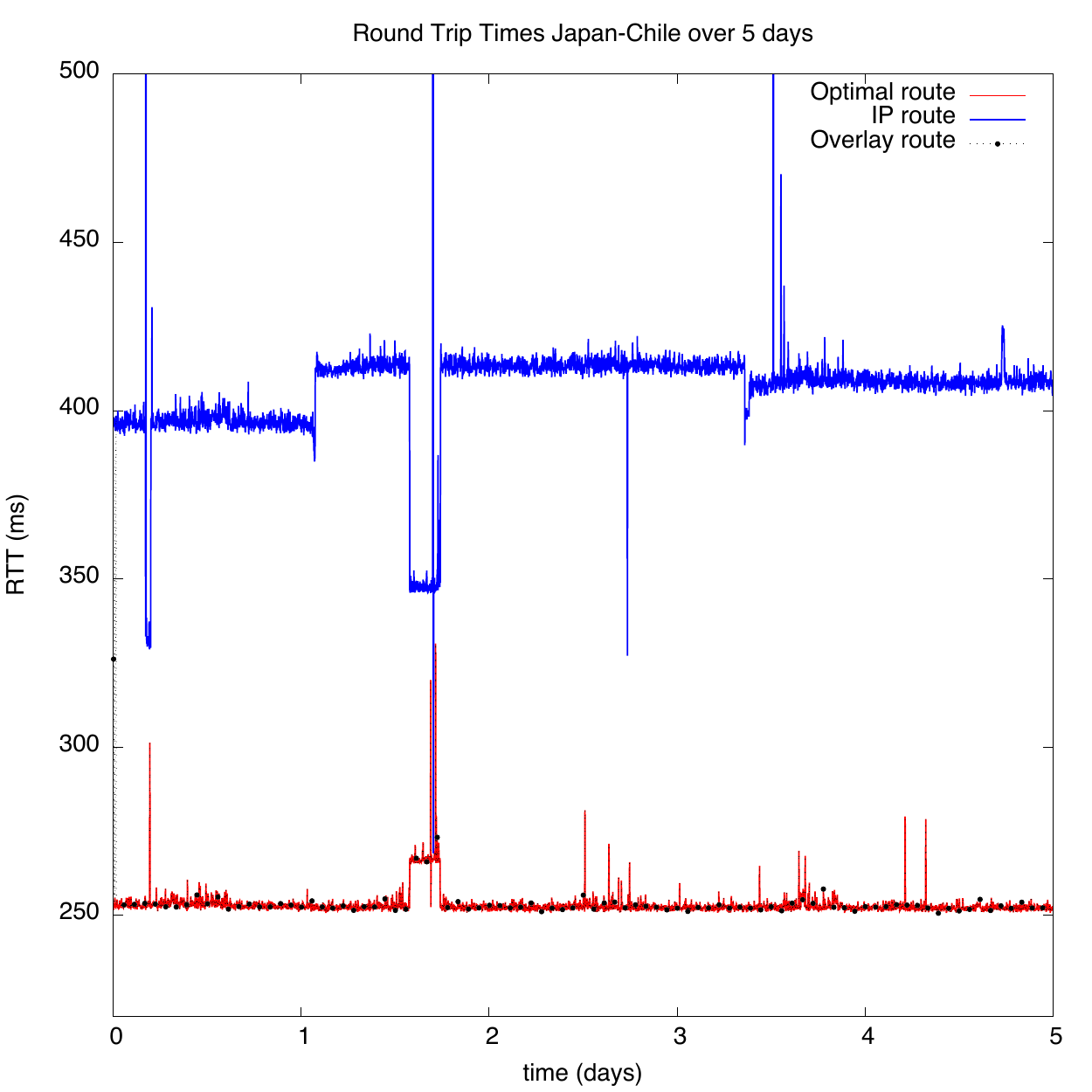}
    \caption{RTD in millliseconds measured for the Japan-Chile connection in an experiment lasting 5 successive days. We see that measured RTD for the the SMART routing policy (in black dots) follows the values of the optimal (i.e. minimum) RTD very closely.}
    \label{fig:RTD-japan-chile-nlnog}
  \end{figure}

\begin{figure}[!htb]
  \centering
    \includegraphics[width=0.8\columnwidth]{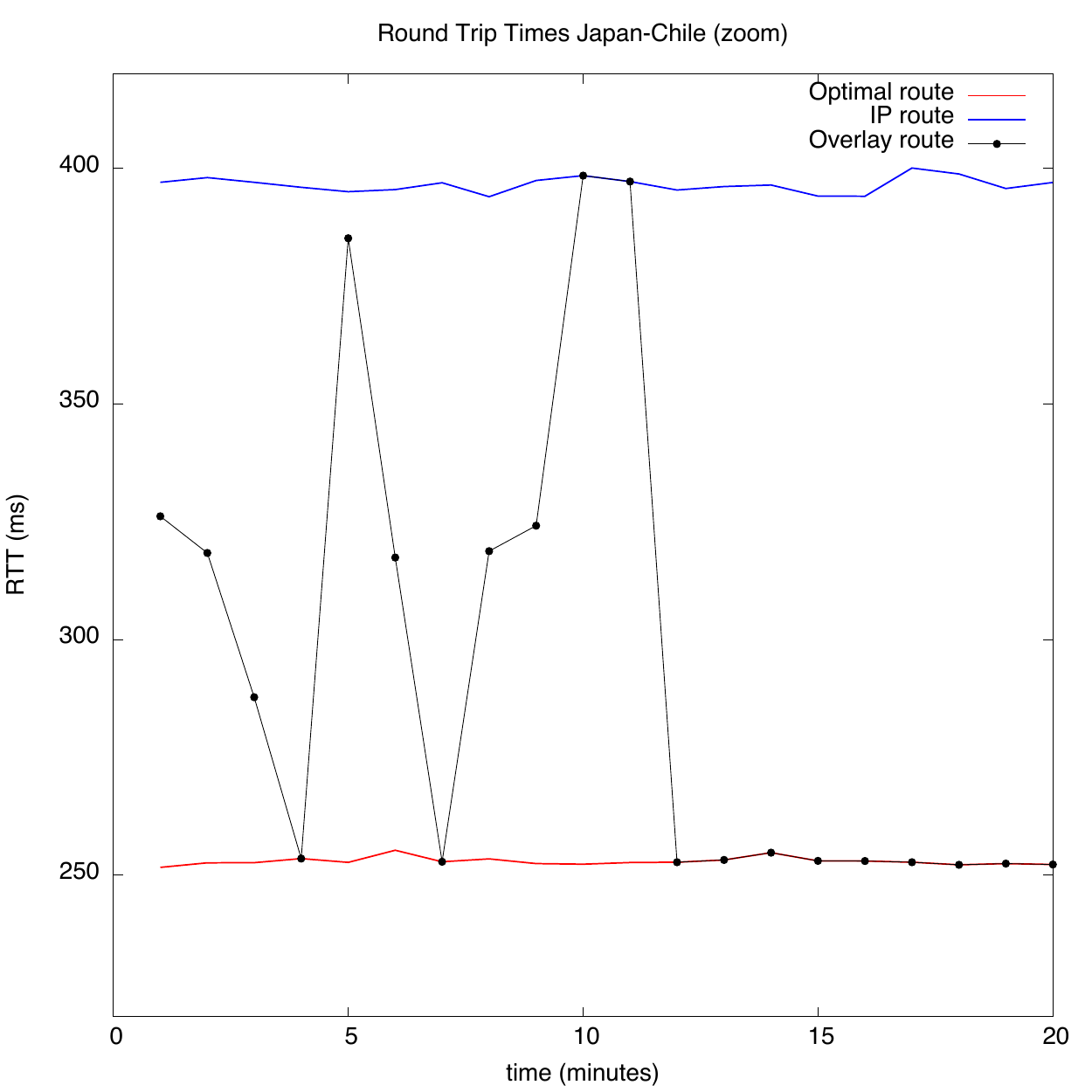}
    \caption{RTD in milliseconds for the Japan-Chile connection for 20 successive 2-minute measurements, over the \underline{first $20$ minutes} of the experiment reported in Figure \ref{fig:RTD-japan-chile-nlnog}. We see that the SMART policy is \underline{learning} right at the beginning, oscillating between the RTD measurements
		for IP (in blue) and the optimum in red. Howeve, after a relatively short time interval, SMART settles to closely following the optimum.}
    \label{fig:RTD-japan-chile-nlnog-zoom}
  \end{figure}

On the other hand, Figure \ref{fig:RTD-japan-chile-nlnog} shows the RTD between the nodes in Japan and Chile over $5$ successive days. The RTD of the direct IP route is about $400$ ms, whereas the RTD of the minimum latency path is about $250$ ms. As can be seen, the RNN-based algorithm learns very quickly which is the minimum latency path and tracks this path until the end of the $5$ days. Figure \ref{fig:RTD-japan-chile-nlnog-zoom} shows the RTDs of the IP route, of the optimal path and of the RNN-based algorithm over the first $20$ minutes. We also notice that it takes only $12$ minutes for our algorithm to learn the optimal route. 

\begin{figure}[!htb]
  \centering
    \includegraphics[width=0.8\columnwidth]{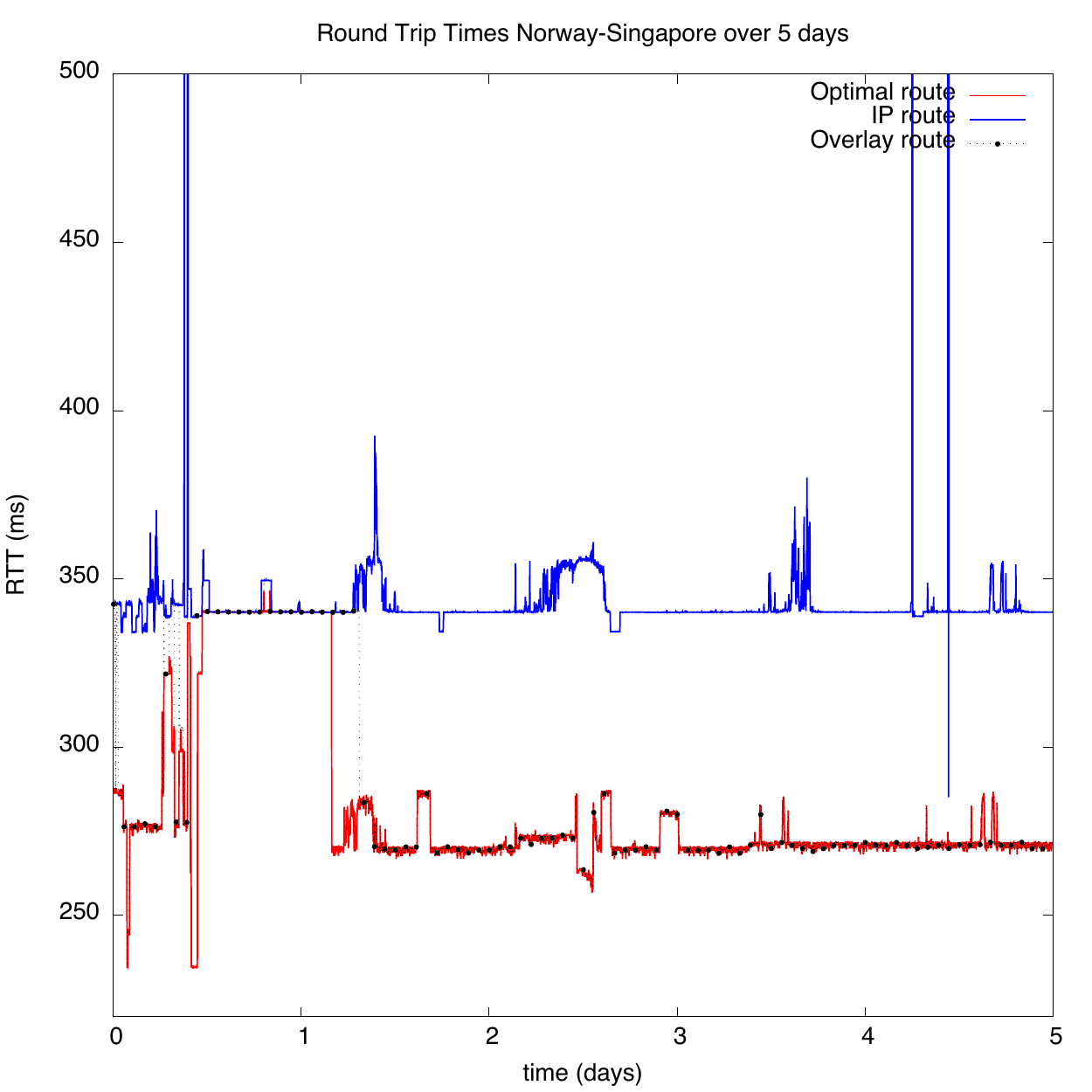}
    \caption{Measured RTD in milliseconds for the Norway-Singapore connection over 5 successive days. Again we observe that the RTD for the SMART scheme (dotted line) closely tracks the measurements for the optimal paths (red), even when they provide the same results as IP routing (blue). }
    \label{fig:RTD-Norway-Singapore-nlnog}
  \end{figure}

\begin{figure}[!htb]
  \centering
    \includegraphics[width=0.8\columnwidth]{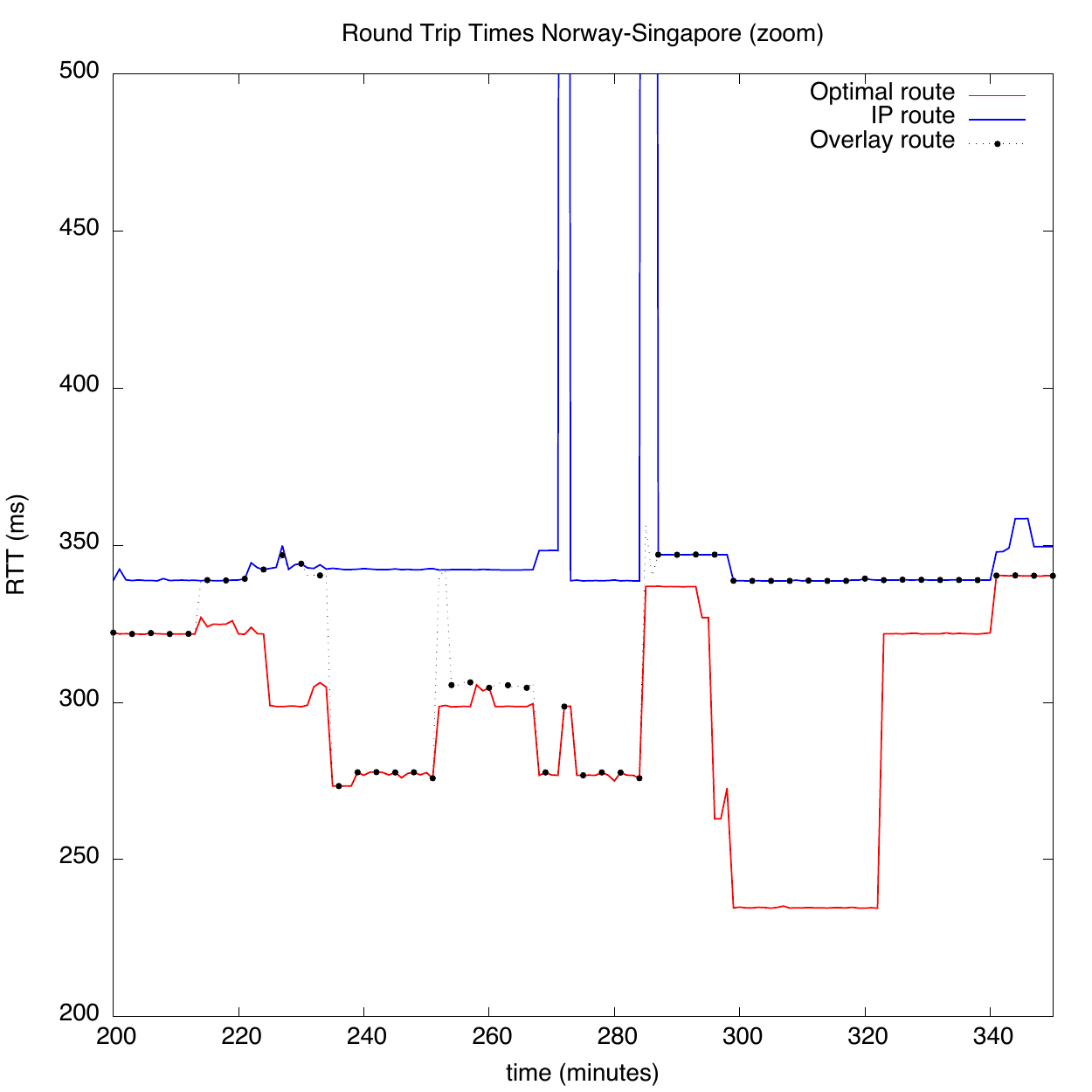}
    \caption{RTD in milliseconds for the Norway-Singapore connection between hours $3$ and $5$ of the experiment of Figure \ref{fig:RTD-Norway-Singapore-nlnog}.
		We notice that in this early stage of the experiment, SMART is unable to achieve the optimum (red) RTD, while from  the previous Figure 
		\ref{fig:RTD-Norway-Singapore-nlnog}, we know that as we enter the later stages of the experiment the optimum is definitely attained thanks to the use of the
		learning offered by the RNN based Reinforcement Learning algorithm.}
    \label{fig:RTD-Norway-Singapore-nlnog-zoom}
  \end{figure}

Figures \ref{fig:RTD-Norway-Singapore-nlnog} and \ref{fig:RTD-Norway-Singapore-nlnog-zoom} provide a similar comparison for the RTD between Norway and Singapore. The RTD of the direct IP route is about $340$ ms, whereas the RTD of the minimum latency path is about $270$ ms. Here again, the RNN-based algorithm learns the minimum latency path very quickly.  However, it does not track the minimum latency path as well as in the previous case, and we can notice some discrepancies between the minimum RTD and the RTD of the overlay during the first day, between hours $3$ and $5$. Figure  \ref{fig:RTD-Norway-Singapore-nlnog-zoom} shows that perturbations can last for a some tens of minutes. Nevertheless, the overlay always provides better performance than what is offered by the IP routing protocols.

\section{Conclusions}
\label{sec:conclusion}

This paper observes that intercontinental IP routes are far from optimal with respect to QoS metrics such as delay
and packet loss, and that they may also be subject to outages, and then develops a {\em Big Data and Machine Learning} approach called SMART, to improve the overall QoS of  Internet connections, while still exploiting the existing IP protocol with path variations offered by an overaly network.
SMART uses an adaptive overlay system that creates a limited modification of IP routes resulting in much lower packet forwarding delays and loss rates. 

The overlays we consider include very few overlay hops, in fact  at most four, and IP is seamlessly used for transporting packets between the overlay nodes. The overlay routing strategy
is inspired from Cognitive Packet Network (CPN) routing, where paths are dynamically selected using a Random Neural Network (RNN) based adaptive learning algorithm that 
exploits smart search and probing in order to select the best paths. In this particular case, each connection explores a small number of alternate paths that are
offered by the overlay network. However, all of this requires constant measurements between the overal nodes resulting in hundreds of thousands of data points being collected in a single day, as well as a fast (RNN based) machine learning algorithm. 

The proposed system has been implemented and tested in an intercontinental overlay network that includes Europe, Asia, North and South America, and Australia, composed of 19 overlay nodes, and we have
observed that with at no more than two overlay nodes in each connection, round trip packet delays are generally very close within a few percent,  to the round trip delays observed with
three or four overlay nodes per connection. We further observe that significant improvements can also be obtained when the RNN based adaptation uses no more than two alternate paths which emerge as the best, as a result of a wider search.

This research can be extended in several directions. An issue to be considered is that of reducing the amount of data that is stored, especially when measurements may have to be stored and exploited concurrently at multiple locations in the network  because paths will typically share overlay nodes and IP network segments. One approach for consideration in future work is 
to resequence the data \cite{Resequencing} so as to drop data items before they enter into the learning algorithm if they have been superseded by fresher and more relevant items. Another important direction is to consider bandwidth optimisation or bandwidth guarantees, in addition to delay minimisation, as is done in recent work
with CPN where applications require asymmetric QoS (delay for upstream and bandwidth for downstream) \cite{Zarina}, as well as energy consumption aspects
in the network as a whole \cite{Energy}. Another question of interest would be to consider the time and energy costs \cite{Time-Energy} of finding users or other resources such as virtual machines or files, in a large overlay network, prior to setting up connections to the appropriate overaly node. 

The results we have obtained have essentially considered paths of at most two overlay hops. This may not be sufficient for some source to destination pairs, especially with complicated
intercontinental patterns of connectivity: for instance Africa can be reached in several different ways, over the Mediterranean Sea, through the Middle East, or along its Atlantic Coast. 

Thus we may need to consider more extensive probing schemes, that use much more data collected at (say)
one or two hour intervals,  which may include probing of all overlay node pairs. Using such data we may be able to  determine the shortest-delay paths between each source to destination pair,
as a means to improve the effectiveness of the adaptive schemes that we propose in this paper.
Indeed, we believe that data driven observation and adaptive control of the large scale Internet is an important area of future work both for researchers and for network operators.

\section{Acknowledgments}
The authors gratefully acknowledge the  support for this research and the funding received from the EC 7th Framework Programme's PANACEA Project (www.panacea-cloud.eu) under Grant Agreement No. 610764.

\bibliographystyle{IEEEtran}
\bibliography{references}  
%

\begin{IEEEbiography}[{\includegraphics[scale=0.5,width=1in,height=1.25in,clip]{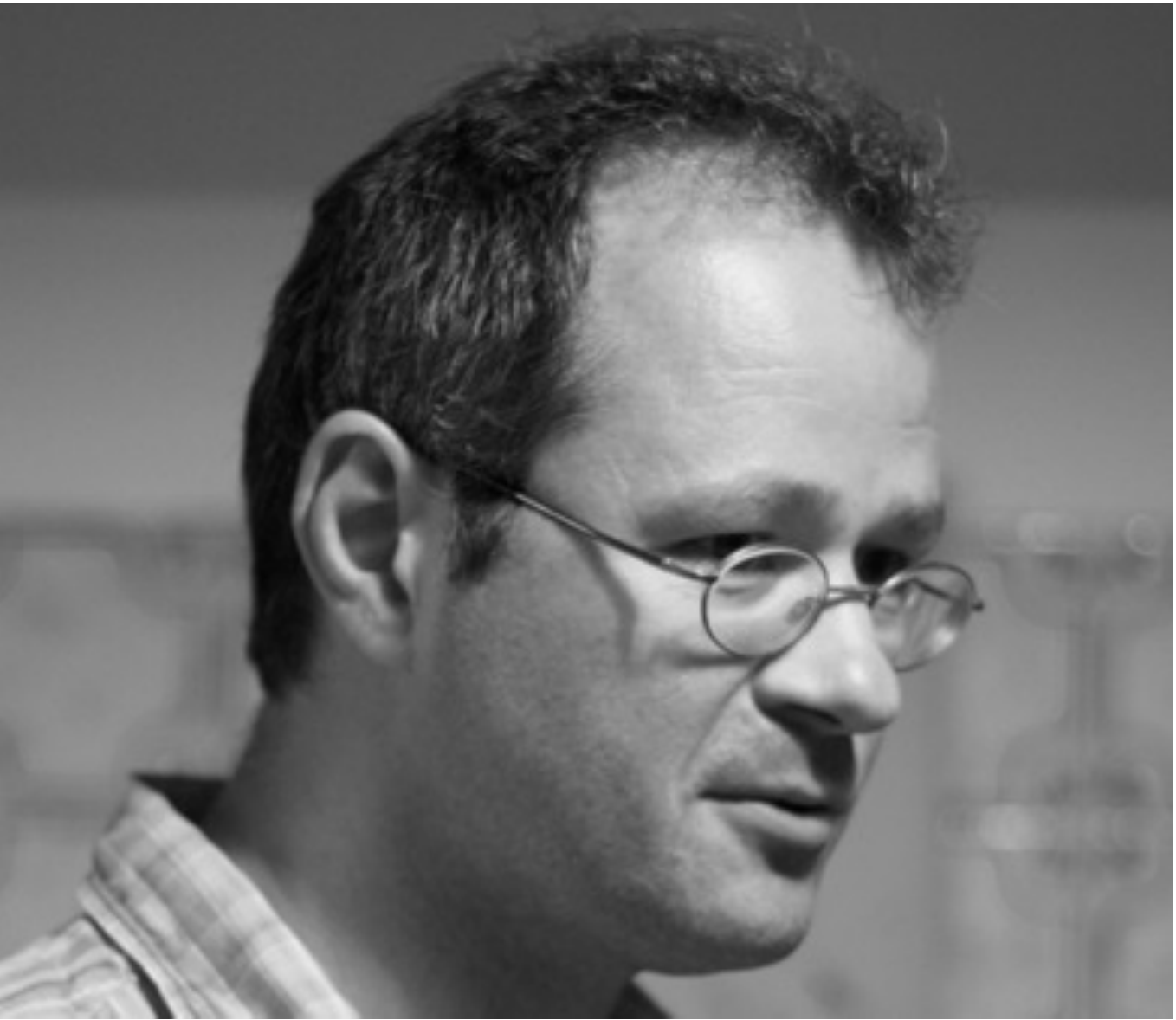}}]%
{Olivier Brun}
is a CNRS research staff member at LAAS-CNRS, Toulouse, France, in the 
SARA group. He graduated from the Institut National des 
T\'el\'ecommunications (INT, Evry, France) and was awarded his PhD 
degree from Universit\'e de Toulouse III (France). His research interests 
include stochastic modeling and performance evaluation of communication 
networks with a special interest in applications of game theory for 
distributed control in networks.
\end{IEEEbiography}
\begin{IEEEbiography}[{\includegraphics[scale=0.5,width=1in,height=1.25in,clip,keepaspectratio]{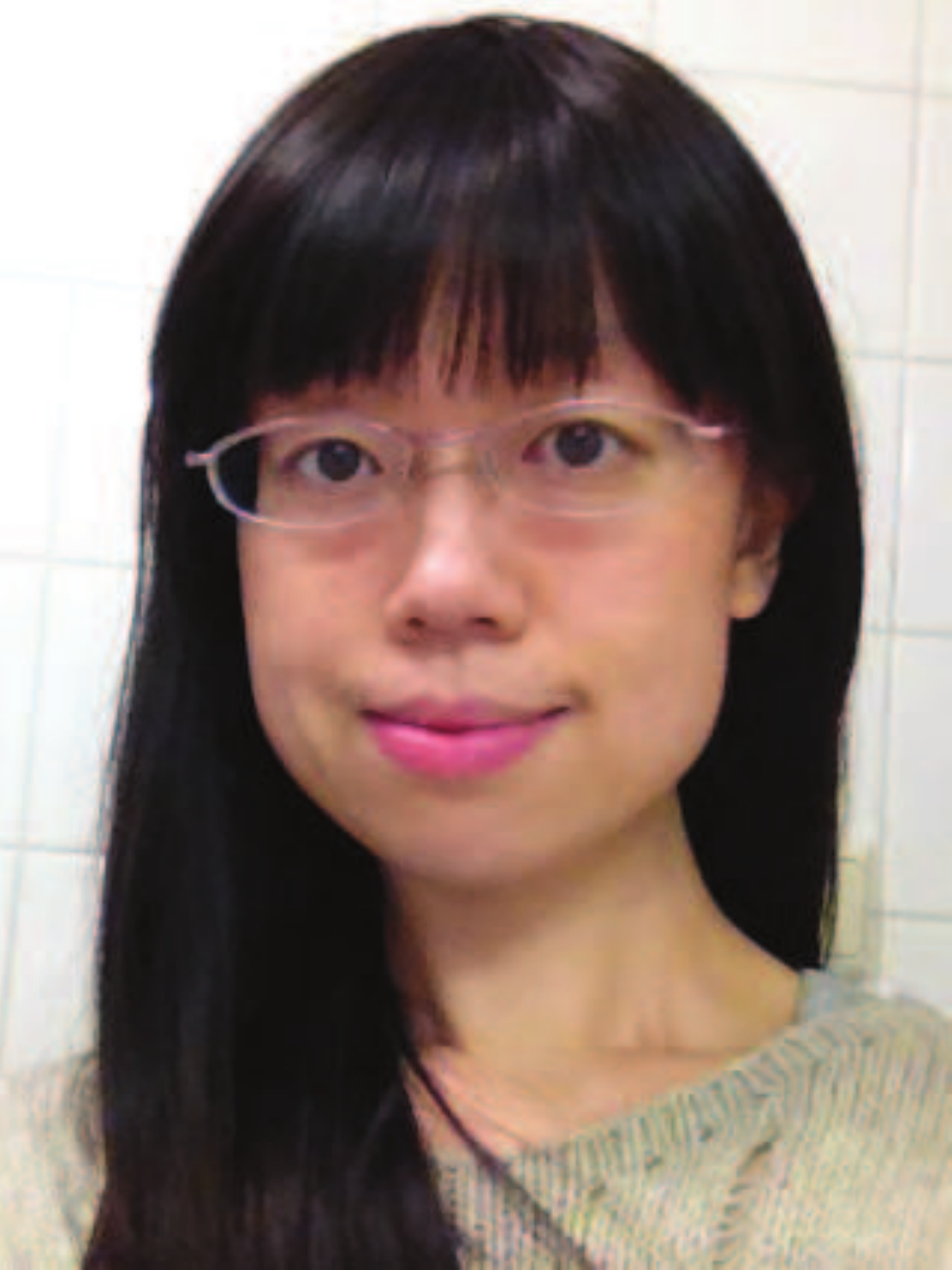}}]%
{Lan Wang}
is currently a PhD student in the Intelligent System and Network group of the Department of Electrical and Electronic Engineering at Imperial College London, working on Quality of Service in Cloud computing systems and computer networks. She obtained her Bachelor's degree in Electronic Engineering from Shandong University and her Master's degree in Communications and Signal Processing from Imperial College London. Before studying at Imperial College, she worked as a software development engineer in integrated telecommunication network management systems in China. She has co-authored several conference papers and two IEEE Transactions papers.
\end{IEEEbiography}
\begin{IEEEbiography}[{\includegraphics[scale=0.5,width=1in,height=1.25in,clip,keepaspectratio]{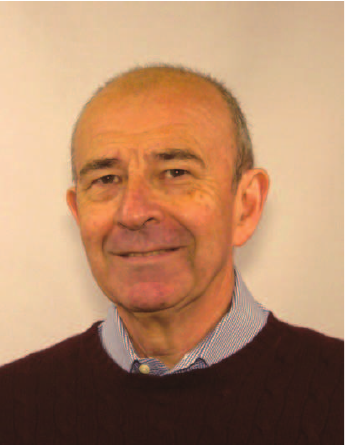}}]%
{Erol Gelenbe}
F'86 FACM is Professor and Head of Intelligent Systems and Networks at Imperial College London, working on energy savings, system security and network Quality of Service. He previously founded the System Performance Evaluation groups at INRIA and French universities, chaired the ECE Department at Duke University, created the School of Electrical Engineering and Computer Science at the University of Central Florida, founded the team that designed the commercial Queueing Network Analysis Package QNAP at INRIA,  and developed the FLEXSIM Flexible Manufacturing System Simulator, the SYCOMORE multiprocessor switch and the XANTHOS fibre optics local area network. In his more theoretically oriented work, he invented mathematical models such as G-Networks for computer and network performance,  and the Random Neural Network to mimic
spiked neuronal behaviour in biological systems as well as learning. Awarded ``Honoris Causa''  doctorates by the University of Li\`{e}ge (Belgium), Universit\`{a} di Roma II (Italy), and Bo\~{g}azi\c{c}i University (Istanbul), he received the ``In Memoriam Dennis Gabor Award'' of the Hungarian Academy of Sciences '13, the Oliver Lodge Medal of the IET (London) '10, the ACM-SIGMETRICS Life-Time Achievement Award '08 and the Grand Prix France Telecom '96. He is a Fellow of the French National Academy of Technologies, the Hungarian, Polish and Turkish Science Academies, and the Royal Academy of Sciences, Arts and Letters of Belgium.
\end{IEEEbiography}

\end{document}